\documentclass[12pt]{iopart}
\usepackage{graphicx}
\usepackage[margin=3cm]{geometry}
\usepackage[english]{babel}
\usepackage{setspace}
\usepackage{tabularx}
\usepackage{color}
\usepackage{amssymb}
\usepackage{xcolor}
\usepackage{soul}
\usepackage[utf8]{inputenc}
\usepackage{xcolor}
\usepackage{ulem}

\begin{document}
\title{Sympatric speciation based on pure assortative mating}
 

\author{R.A. Caetano}
\address{Departamento de F\'isica ,
 Universidade Federal do Paran\'a, Caixa Postal 19044, 81531-990, Curitiba, PR, Brazil}
\ead{caetano@fisica.ufpr.br}
\vspace{10pt}

\author{Sergio Sanch\'ez}
\address{Departamento de F\'isica ,
 Universidade Federal do Paran\'a, Caixa Postal 19044, 81531-990, Curitiba, PR, Brazil}
\ead{ssanchez@fisica.ufpr.br}
\vspace{10pt}

\author{Carolina L. N. Costa}
\address{Instituto de Biologia ,
Universidade Estadual de Campinas, 13083-859, Campinas, SP, Brazil}
\ead{lemes.carol@gmail.com}
\vspace{10pt}

\author{Marcus Aloizio Martinez de Aguiar}
\address{Instituto de F\'{\i}sica `Gleb Wataghin', 
Universidade Estadual de Campinas, 13083-970, Campinas, SP, Brazil}
\ead{aguiar@ifi.unicamp.br}
\vspace{10pt}
\vspace{0.5cm}






\begin{abstract}
Although geographic isolation has been shown to play a key role in promoting reproductive isolation, it is now believed that speciation can also happen in sympatry and with considerable gene flow. Here we  present a model of sympatric speciation based on assortative mating that does not require a genetic threshold for reproduction, i.e., that does not directly associate genetic differences between individuals with reproductive incompatibilities. In the model individuals mate with the most similar partner in their pool of potential mates, irrespective of how dissimilar it might be. We show that assortativity alone can lead to the formation of clusters of genetically similar individuals. The absence of a minimal genetic similarity for mating implies the constant generation of hybrids and brings up the old problem of species definition. Here, we define species based on clustering of genetically similar individuals but allowing genetic flow among different species. We show that the results obtained with the present model are in good agreement with empirical data, in which different species can still reproduce and generate hybrids.
\end{abstract}
\section {Introduction}
\label{intro}

One of the main goals of evolutionary biology is to uncover the processes responsible for the origin
and maintenance of biodiversity. However, the very concept of species has been the subject of an
endless discussion in evolutionary theory
\cite{mayr1942systematics,mallet1995species,paterson1985recognition,templeton1989meaning,
van1976ecological,wiley1978evolutionary,cracraft1987species,queiroz1988phylogenetic,shaw2001genealogical, baker2006speciation}. While useful from a theoretical point of view, the famous Biological Species Concept (BSC), based on reproductive isolation, is very hard to test in practice and cannot deal adequately with hybridizations \cite{harrison2014hybridization}.  According to the BSC, a species is a group of organisms that can reproduce with one another and that are reproductively isolated from other such groups. When the mating pattern of the individuals is assortative, i.e., when mating occurs preferentially between individuals with similar phenotypes, the  gene flow between species is limited, helping to keep them separated \cite{kirk2013,jiang2013assortative}. However, individuals from different species do eventually mate and produce viable offspring, posing a challenge to the BSC definition. This is the case of the butterfly {\it Heliconius cydno cydnides} and {\it Heliconius cydno weymeri} that live in sympatry (same geographic area) in the Cauca Valley (Colombia). It has been shown that {\it Heliconius cydno cydnides} males court and produce viable offsprings with {\it Heliconius cydno cydnides} and {\it Heliconius cydno weymeri} females, but they have a strong preference for females of the same species \cite{arias2012}. Perhaps the most celebrated case of in which mating is assortative but interbreeding produces fertile hybrids is the cichlid fish in the African lakes \cite{kocher2004adaptive}. Lakes Malawi and Victoria, in particular, exhibit an exuberant fauna of haplochromine cichlids. There are convincing evidences of female choice based on male coloration in both artificial conditions and natural environments \cite{dieckmann2004adaptive}. Nevertheless, under certain conditions, such as inappropriate lighting environment, mate choice becomes random and viable offspring are produced \cite{seehausen1997cichlid}. Assortative mating is believed to be the main genetic barrier between the butterfly species and it is also invoked to explain the explosive radiation of cichlid fish in African lakes \cite{kocher2004adaptive}.  

The incorporation of assortativity in speciation models is a way to understand the dynamics of speciation in systems such
as {\it Heliconius} and African cichlid fish, where species have possibly diversified in sympatry with assortment playing an important role as a source of reproductive isolation \cite{kirkpatrick2002speciation,coyne2004speciation,gavrilets2004fitness}. Models with assortative mating can be distinguished between one- and two-trait models \cite{shpak,fry2003multilocus}. These two kinds of models differ on whether the set of loci under viability selection and the loci involved in assortative mating are the same or not \cite{de2008model}. Selectively neutral one-trait models have been shown not to lead to speciation due to the stabilization of the sexual selection generated by assortative mating \cite{kirkpatrick2000reinforcement,kirkpatrick2004sexual,bahar}. The combination of selection and assortativeness, however, can be very effective at eliminating intermediary genotypes, resulting in sympatric speciation \cite{10.2307/2408632, 10.2307/2410209}. Alternatively, assortative mating combined with spatial structure and mating with nearby individuals  can also lead to the formation of local groups of similar individuals that eventually break into species \cite{Gavrilets2000b,Hoelzer2008,Fitzpatrick2009,DeAguiar2009}. Two-trait models, which separate the trait under selection – such as body size or eye color – and the trait under assortative mating, have showed that, if linkage is tight and assortment is strong, reproductive isolation can emerge \cite{Dieckmann1999,kondrashov1999interactions}. 

An important model of assortative mating was proposed by  Derrida and Higgs (DH) \cite{higgs1991stochastic}, which demonstrated the possibility of speciation in sympatric populations involving no other selective forces or spatial structure. One key aspect of the model is the approximation of infinitely large genomes \cite{de_Aguiar_2017}. In this model assortative mating is controlled by a measure of genetic similarity between individuals. An incompatibility threshold $q_{min}$ is introduced in such a way that reproduction becomes impossible between individuals $i$ and $j$ whose similarity $q_{ij}$ is smaller than $q_{min}$. The incompatibility threshold allows the definition of species as groups of individuals reproductively isolated from those in different groups, similar to the BSC. This definition is very convenient for theoretical models and has been adopted by many authors \cite{baker2006speciation,	martins2013evolution,baptestini2013role,schneider2016diploid,higgs1991stochastic,de_Aguiar_2017}. Given a similarity threshold $q_{min}$, it is easy to design an algorithm that separates the population into clusters in such way that individuals belonging to different clusters have genetic similarity smaller that $q_{min}$. Within clusters, on the other hand, individuals will have at least one partner  whose genetic similarity is larger than $q_{min}$. The process is similar to finding components of a network where nodes (individuals) are connected only if their similarity is larger than $q_{min}$.  However, in real cases of incipient speciation such threshold may not exist and the populations will not separate in perfectly isolated groups, as illustrated by {\it Heliconius} and African cichlid fish systems, in which different species can still reproduce and generate hybrids.

Here we propose a model of speciation based on pure assortativity, where no spatial structure  or
similarity threshold to reproduction are imposed. The evolutionary dynamics is based  on the model
by Derrida and Higgs \cite{higgs1991stochastic}, with the difference that individuals will mate with the
most similar individual in their pool of potential partners, characterizing positive assortative
mating, but with no restrictions imposed by a similarity threshold.  We show that assortative mating
alone can lead to the formation of groups of genetically similar individuals that will reproduce
preferentially with others in the same group, but might occasionally choose individuals from other groups
as mating partners, generating viable hybrids. One important feature of the model is that it brings
back the  discussion of species definition: the clusters that form are not reproductively isolated
from each other and are constantly, though rarely, generating hybrids. This recovers the case of
African cichlids and other incipient species where mate choice does not seem to be related to
genetic incompatibilities. The model provides a good representation of speciation by assortative
mating in a genetically neutral scenario.




\section {Methods}

 \subsection{Assortative Mating} 
 
 Our model is a modification of the model proposed by Derrida and Higgs (DH) \cite{higgs1991stochastic}, in which a
population of M haploid individuals evolves by sexual reproduction with a genetic threshold determining
mating compatibility. Each individual is characterized by a genome represented by a
binary sequence of size $B$, $\{S_{1}^{\alpha},S_{2}^{\alpha},\dots,S_{B}^{\alpha}\}$, where $S_{i}^{\alpha}$ is the
$i^{th}$ gene of the individual $\alpha$ and can assume the values $\pm 1$. The population is characterized by a $M \times M$ matrix $Q$ measuring the degree of genetic similarity between pairs of individuals. The elements of $Q$ are given by:

\begin{equation}
\label{q}
q^{\alpha\beta}=\frac{1}{B}\sum_{i=1}^{B}S_{i}^{\alpha}S_{i}^{\beta}.
\end{equation}
Genetically identical individuals have $q^{\alpha\beta}=1$ whereas individuals with randomly assigned genes have
$q^{\alpha\beta}$ close to zero. Also, since every individual is identical to itself, $q^{\alpha\alpha}\equiv 1$ at all times. All results in
this paper are obtained in the limit of infinitely large genomes, $B \rightarrow \infty$ as described below.

In the original DH model the population at each generation is obtained from the previous one by
sexual reproduction of parents with a minimal genetic similarity threshold $q_{min}$, as follows: an
individual is randomly chosen from the population to be the first parent, $P_{1}$. Then a second
parent, $P_2$ is selected from the remaining $M-1$ individuals. If $P_1$ and $P_2$ are sufficiently
similar, i.e., if $q^{P_1,P_2} \geq q_{min}$ an offspring is produced by combining the genomes of
the parents. If, however, $q^{P_1,P_2} < q_{min}$, the pair is considered incompatible and another
individual is randomly chosen for the role of $P_2$. If the similarity condition cannot be met by
any member of the population, another first parent $P_1$ is chosen and the process repeated until a
compatible mating pair is found. The genome of the offspring is inherited, gene by gene, from the
parents' genomes with equal probability. There is also a small probability, $\mu$, that a mutation
occurs in each gene of the offspring. The entire process is repeated until $M$ offspring have been
produced, forming the next generation.

The idea behind the hypothesis of a minimum similarity for mating is reminiscent of the Batenson-Dobzhansky-Muller 
model \cite{bateson1909heredity,dobzhansky1937genetics,muller1939reversibility,muller1942isolating} where alleles at different loci of diploid individuals might be incompatible. Here each difference between haploid
genomes reduces the chances of producing a viable offspring by introducing some degree of
incompatibility. As the number of differences increases the chances of a successful reproduction
decreases. A simple way to deal with this idea is by assuming that individuals become totally
incompatible if the number of genetic differences exceeds a threshold but are perfectly compatible
below that value. However, such a sharp threshold is a simplification of the real mating
dynamics as genetic incompatibilities might not exist at all, as in the case of cichlids \cite{seehausen1997cichlid}.

In the present model the condition of minimum similarity for mating is completely removed. Instead,
after a first parent $P_1$ has been randomly chosen, a pool $R$, with $N$ individuals is randomly
selected from the remaining population: $R=\{R_{1}, R_{2}, \dots, R_{N}\}$. The second parent is now
chosen from the pool as the individual with the highest degree of similarity with $P_{1}$. Thus,
$P_{2}=R_{i}$, such that: $q^{P_{1}R_{i}} = max(q^{P_{1}R_{j}})$, with $j=1,2,\dots,N$. Mating is,
therefore, purely assortative and relies on the power of the first parent to choose its mate from
the pool, with no restriction regarding genetic distance. An offspring is always produced by
combining the genomes of $P_1$ and $P_2$, no matter how different they might be. Like in the
original DH model, parents' genomes are combined gene by gene with a mutation rate $\mu$ per gene and
the process is repeated until $M$ offspring have been generated.

The model we propose here can be interpreted as a type of  {\it best-of-n} model \cite{alexey,burrow}. In this case, a female evaluates $n$ randomly selected males and chooses the one closest to her preferred phenotype . It has been show that {\it best-of-n} models may lead to speciation, however not in the neutral context as in the present work \cite{Servedio8113}

In order to get some insight on how the overlap matrix evolves through time, we follow DH and
consider first the simplest case of asexual reproduction, in which each individual $\alpha$ has a
single parent $P(\alpha)$. The allele $S_{i}^{\alpha}$ has probability
$\frac{1}{2}(1+e^{-2\mu})$ of inheriting the allele $S_{i}^{P(\alpha)}$ without
mutating and probability $\frac{1}{2}(1-e^{-2\mu})$ of undergoing a mutation
and assuming the value $-S_{i}^{P(\alpha)}$. Under these conditions, the expected value of 
$q^{\alpha\beta}$ is:

\begin{equation}
\label{media}
E(q^{\alpha\beta})=e^{-4\mu}q^{P(\alpha)P(\beta)}
\end{equation}

In the case of sexual reproduction the individuals $\alpha$ and $\beta$ have two parents each:
$P_{1}(\alpha)$, $P_{2}(\alpha)$ and $P_{1}(\beta)$, $P_{2}(\beta)$, respectively. On average, half
of the alleles are inherited from $P_{1}$  and the other half from $P_{2}$. Thus, the expected value
of $q^{\alpha\beta}$ is:

\begin{equation}
\label{qsex}
E(q^{\alpha\beta})=\frac{e^{-4\mu}}{4}\left[q^{P_{1}(\alpha)P_{1}(\beta)}+q^{P_{2}(\alpha)P_{1}(\beta)}+q^{P_{1}(\alpha)P_{2}(\beta)}
+q^{P_{2}(\alpha)P_{2}(\beta)}\right]
\end{equation}

The above expressions are exact for infinitely large genomes ($B \rightarrow\infty$) and the
dynamics of the population can be obtained by simply updating the similarity matrix. Therefore,
given the population at generation $t$ and similarity matrix $Q$, the next generation is constructed
by producing $M$ offspring with parents at generation $t$ as described above. Equation \ref{qsex} is
then used to update $Q$. If mating is not restricted by similarity and if the pool size is $N=1$, it
can be shown that all elements $q^{\alpha\beta}$ of the similarity matrix converge to the value
$q_{conv}\approx\frac{1}{(1+4\mu M)}$ \cite{higgs1991stochastic}. The population will break into
species if a threshold value $q_{min} > q_{conv}$ is introduced \cite{higgs1991stochastic}. Here we
show that the population still breaks into species if the minimum similarity requirement is
eliminated but pool sizes for choosing the second parent are sufficiently large.

\subsection{Species Definition}
\label{specie}

In the DH model of speciation two individuals $i$ and $j$ cannot reproduce if $q^{ij} < q_{min}$. This genetic
compatibility threshold at the individual level can be extended to the species level to define reproductive isolation
between groups of individuals, as in the biological species concept. In the DH model a species can be identified as a
group of individuals reproductively isolated from all others outside the group by the genetic threshold $q_{min}$. Not
all members of the group have to be able to mate with each other, but could maintain an indirect gene flow through an
intermediary individual. Therefore, if $q^{ij} > q_{min}$ and $q^{jk} > q_{min}$ but $q^{ik} < q_{min}$, individuals $i$,$j$
and $k$ will belong to the same species, owing to the ongoing gene flow between them.

In the present model of pure assortative mating no such threshold exists and, as discussed before, the concept of
species is fuzzy and relies on the identification of clusters of similar individuals. In order to do that we are going
to define an auxiliary  threshold value $q_0$ that will play the role of $q_{min}$ in our model for the sake of species
definition. This value will be calculated at a specific generation, when the population dynamics reaches a state of
equilibrium, and will be used to organize the population into species in all generations, before and after the moment of
calculation. Our definition of $q_0$ will be based on the clustering method proposed in \cite{ng2002spectral} and 
\cite{von2007tutorial} which uses the eigenvalues of the unnormalized Laplacian matrix defined as 
\begin{equation}
\label{laplacian}
L=D-Q.
\end{equation}
Here $Q$ is the similarity matrix whose elements are given by equation (\ref{q}) and $D$ is the {\it degree matrix},
whose elements are given by: 
\begin{equation}
\label{d}
d_{i,j}=\left\{\begin{array}{rc} 0, & \mbox{if}\ i\ne j,\\
\sum_{k=1}^{M} q^{ik}, & \mbox{if}\ i = j\end{array}\right.
\end{equation}

One important property of $L$ is that for every vector $f \in \mathbb{R}^{N}$
\begin{equation}
\label{prop1}
f^{T}Lf=\frac{1}{2}\sum_{i,j=1}^{M}q^{ij}(f_{i}-f{j})^{2}.
\end{equation}
If all elements $q^{ij}$ are positive, this would imply that $L$ has $M$ non-negative eigenvalues
and that 0 is an eigenvalue associated with the constant eigenvector. Although the $q^{ij}$'s 
can be negative (for example if $S^i_k=1$ and $S^j_k=-1$ for all $k$, $q^{ij}=-1$)  they are always positive in the 
simulations, as we start with $q^{ij} = 1$ and evolution keeps them positive at all times. 
Thus, it is possible to establish the following proposition: if $k$ is the multiplicity of the eigenvalue 0 of $L$, then the
graph represented by $Q$ has $k$ disconnected components \cite{von2007tutorial}. Based on these
properties we propose the following procedure to determine $q_{0}$: we define a new similarity
matrix $Q(s)$ whose elements are given by: 
\begin{equation}
\label{newq}
q^{ij}(s)=\left\{\begin{array}{rc} 0, & \mbox{if}\ q^{ij} < s\\
q^{ij}, & \mbox{if}\ q^{ij} > s \end{array}\right. .
\end{equation}
In terms of graph theory this is equivalent to consider vertices with connection smaller than $s$ to
be  disconnected. From $Q(s)$ one obtains $D(s)$ and $L(s)$. Therefore, following the number $k$ of
eigenvalues 0 (the multiplicity of the eigenvalue 0) as a function of $s$, it is possible to find an
appropriate value of this parameter that breaks the population into reasonably disconnected clusters
and use this value as the definition of $q_0$. In other words, a cluster is defined as a group of
individuals (vertices) that have larger genetic similarity  among themselves than with individuals
belonged to other clusters. Thus, the critical value $q_{0}$ is the lowest connection between
vertices belonging to the same clusters. Disconnecting vertices with similarity larger than $q_{0}$
would produce a large number to totally disconnected individuals, causing the number of clusters to
increase rapidly with $s$.

\begin{figure}
	\centering
	\includegraphics[width=15cm]{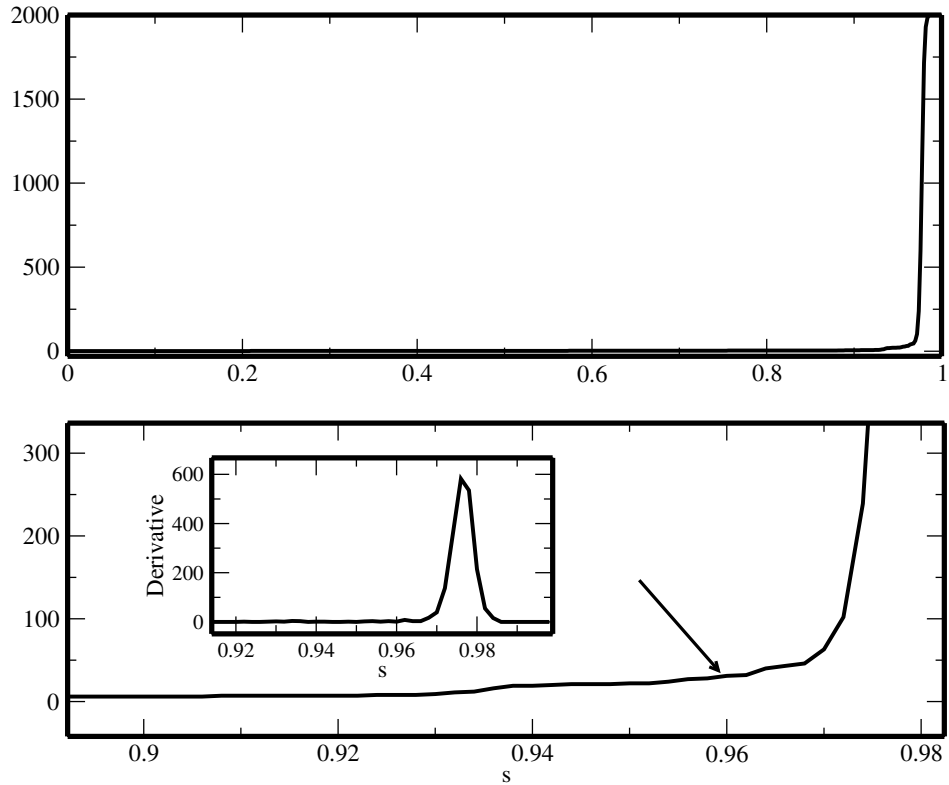}
	\caption{\label{fig:figure1} (upper panel) Number of eigenvalues $k(s)$ smaller than 10  for a population of $M=2000$, pool size $N=100$, 
		after 10000 generations. (bottom panel) Magnification of the region around $s= 0.95$. The inset shows the derivative $dk/ds$. }
\end{figure}

As an example, consider the hypothetical case of a population with $M$ individuals that is clearly
split into two groups. Suppose the first group has $M_{1}$ individuals and that the minimum
similarity between all its pairs is equal to $s_{1}$;  the second group, with $M_{2}$ individuals,
has minimum similarity between its pairs equal to $s_{2}$. We further assume that the minimum
similarity between individuals belonging to different groups is $s_{3}$. Clearly $s_{3}<s_{1}$ and
$s_{3}<s_{2}$ and, without losing generality, we assume $s_{2}<s_{1}$. In this case the behavior of
$k(s)$ is as follows: For $ 0 \leq s \leq s_3$, $k(s)=1$ since all pairs have similarity larger than
$s_3$.  For $s_{3}<s<s_{2}$, $k(s) = 2$, the two groups were disconnected and the multiplicity of
the eigenvalue zero is 2. Finally, for $s_2 < s < s_1,$ individuals belonging to the second cluster
gradually become disconnected from the cluster and $k$ starts to increase continuously. We may thus
consider that $q_0 = s_2$ is the appropriate value of s that splits the population into components
without fragmenting it into tiny clusters with few individuals each. The value, although still
subjected to a degree of uncertainty, can be identified by the presence of a plateau in the plot of
$k(s)$, before a sharp increase. It is important to stress that, different from the BSC, $q_{0}$
does not play a role in mating. In the present model, it is only a convenient parameter to define
species. However, the strength of the present model is the possibility of mating between individuals
with genetic similarity smaller than $q_{0}$.

One representative example from our simulations is shown in Figure \ref{fig:figure1} where we
analyzed  a population with 2000 individuals with pool size of 100 at the 10000-th generation.  In
this case the eigenvalues of the Laplacian matrix range typically from 0 to 3000 and we counted as $k(s)$, or
"close to zero", those below 10. The upper panel of Figure \ref{fig:figure1} shows
$k(s)$ and the bottom panel shows a zoom around $s=0.96$ with the derivative $dk/ds$ in the inset. From this
figure we defined $q_{0} = 0.96$ and use it as a criterion  to define species. We acknowledge that there
is a degree of subjectivity in this procedure. Different choices of $s$ would change the counting of species, but not the clustering 
itself.

\section{Results}

In this section we show results for a population with $M=2000$ individuals and $\mu=1/8000$. The simulation starts with
all individuals genetically identical ($q^{\alpha\beta}=1$ for all $\alpha$ and $\beta$) and the expected equilibrium value of the similarity for the
case of unit pool ($N=1$) is $q_{conv}=1/2$.  

\subsection{Species and similarities}

\begin{figure}
\centering
\includegraphics[width=15cm]{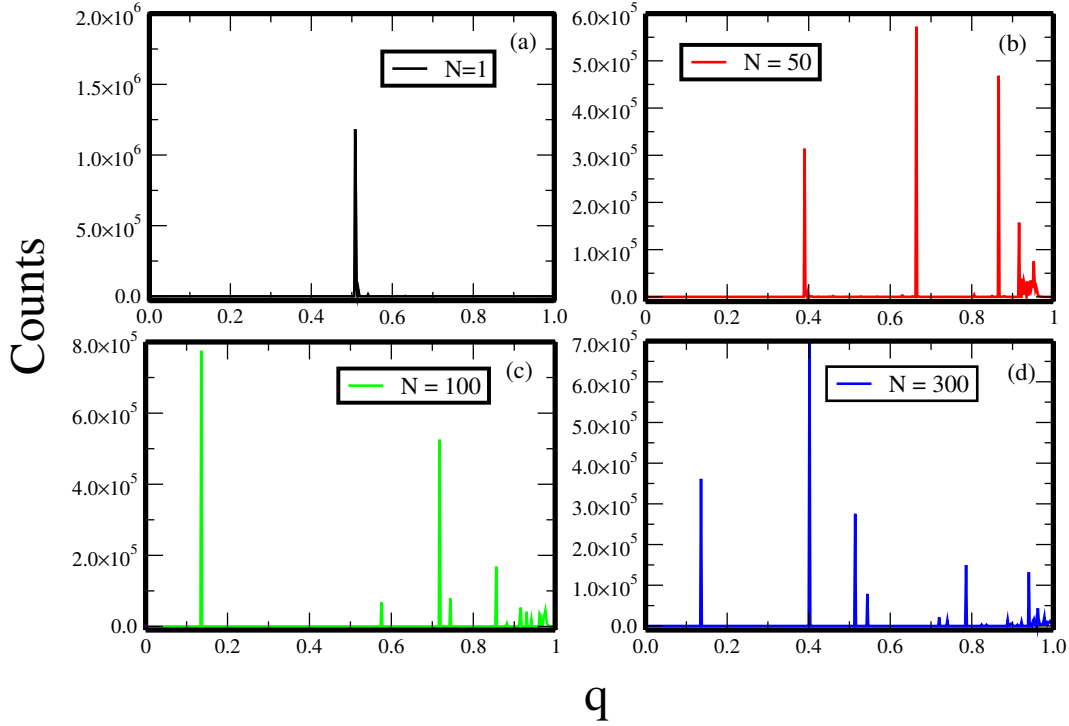}
 \caption{\label{fig:figure2} Distribution of non-diagonal elements of the similarity matrix for a population of $M=2000$
 	individuals and pool sizes (a) 1; (b) 50; (c) 100 and (d) 300 at the generation $T=4000$. Panel (a) is the special 
 	case in which our model recovers the DH model. In all cases we have used $\mu = 1/8000$. }
\end{figure}

\begin{figure}
	\centering
	\includegraphics[width=15cm]{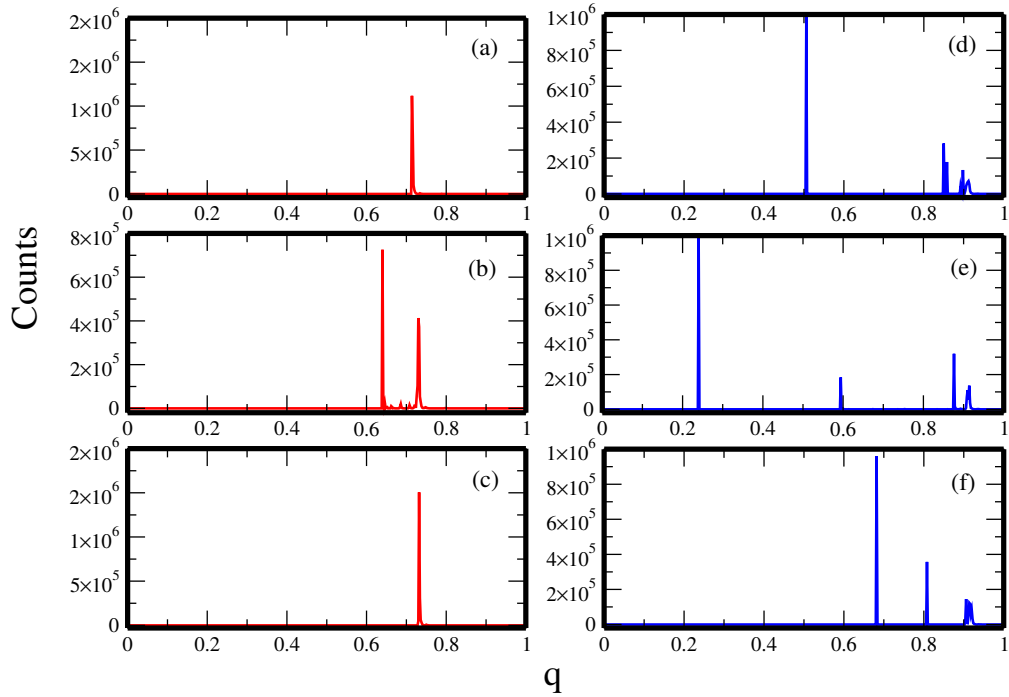}
	\caption{\label{fig:figure3} Distribution of non-diagonal elements of the similarity matrix of a population of $M=2000$
		individuals for pool sizes 10 (red, left panel) and  30  (blue, right panel). In both cases $\mu = 1/8000$.  
		From top to bottom the times are 4000, 5500 and 6000.	}
\end{figure}

Figure \ref{fig:figure2}a shows the histogram of $q^{\alpha\beta}$ between
all pairs of individuals after 4000 generations for random mating ($N=1$) displaying the peak at $q=0.5$ as expected.
Figures \ref{fig:figure2} b,c and d show similar results for pool sizes $N=50$, $100$ and $N=300$ respectively. The
histograms exhibit peaks at large values of $q$, such as between 0.9 and 1.0, indicating that there are clusters of
individuals with more than $90\%$ of genetic similarity. The peaks at lower values of $q$, on the other hand, 
correspond of pairs of individuals in different clusters, which is a signature of speciation.  Figure  \ref{fig:figure3} shows 
results for small pools of sizes 10 and 30.

\begin{figure}
	\centering
	\includegraphics[width=15cm]{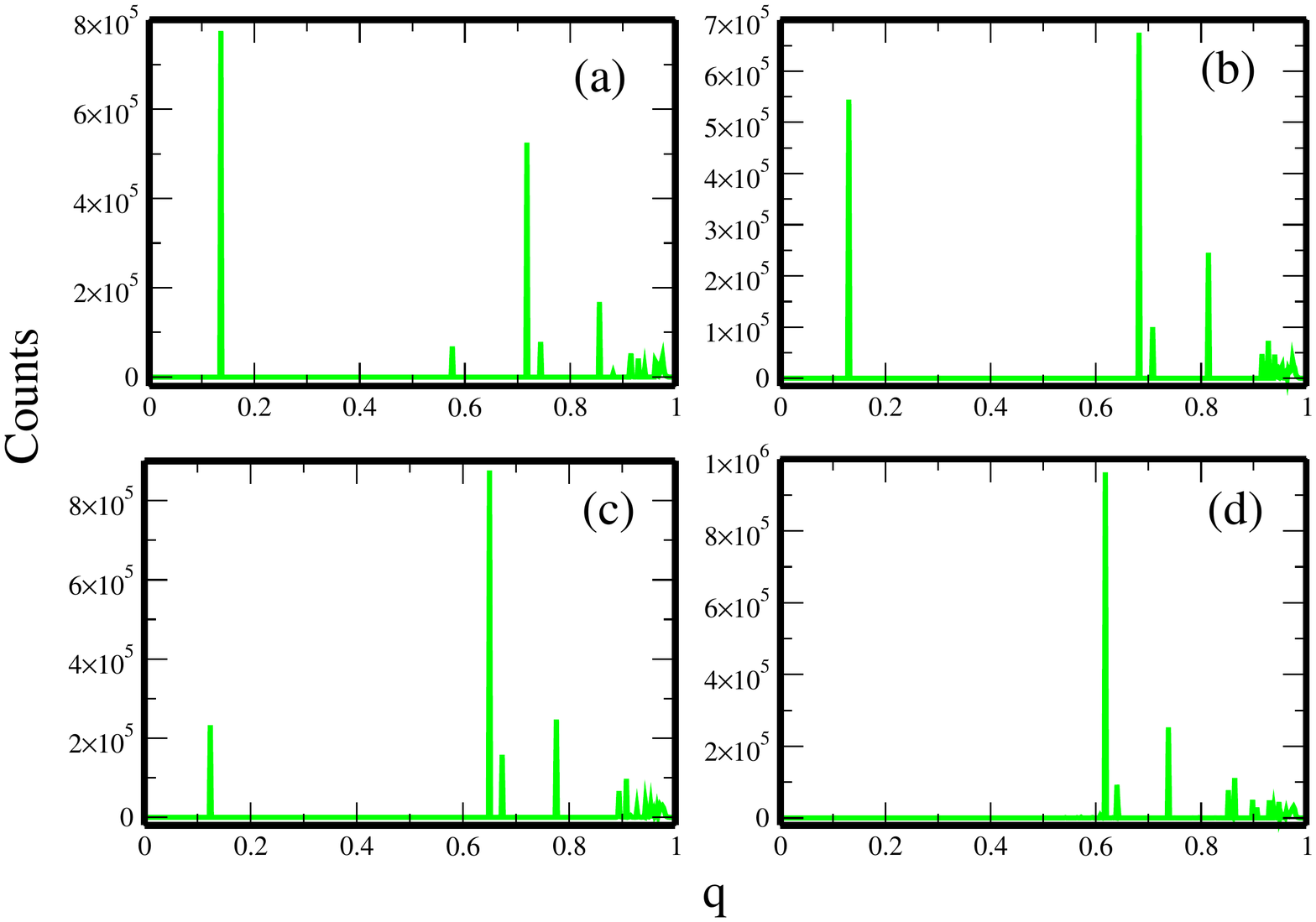}
	\caption{\label{fig:figure4} Time evolution of the distribution of non-diagonal elements of the similarity matrix for pool size 
		$N=100$ and times 4000, 4100, 4200 and 4300.}
\end{figure}

Similarly to the DH model \cite{higgs1991stochastic} the peaks move to the left as time increases because of genetic
drift and eventually disappear as species go extinct by fluctuation in the number of individuals. Figure
\ref{fig:figure4} shows histograms of $q^{\alpha\beta}$ for $N=100$ at the generations 4000 (Figure \ref{fig:figure4}a), 4100 (Figure \ref{fig:figure4}b), 4200 (Figure \ref{fig:figure4}c) and 4300 (Figure \ref{fig:figure4}d).
The peak close to $q=0.1$, for instance,  moves slowly to left  with decreasing amplitude, until it disappears
completely  (Figure \ref{fig:figure4}d). 

\subsection{Number of species}

\begin{figure}
\centering
\includegraphics[width=15cm]{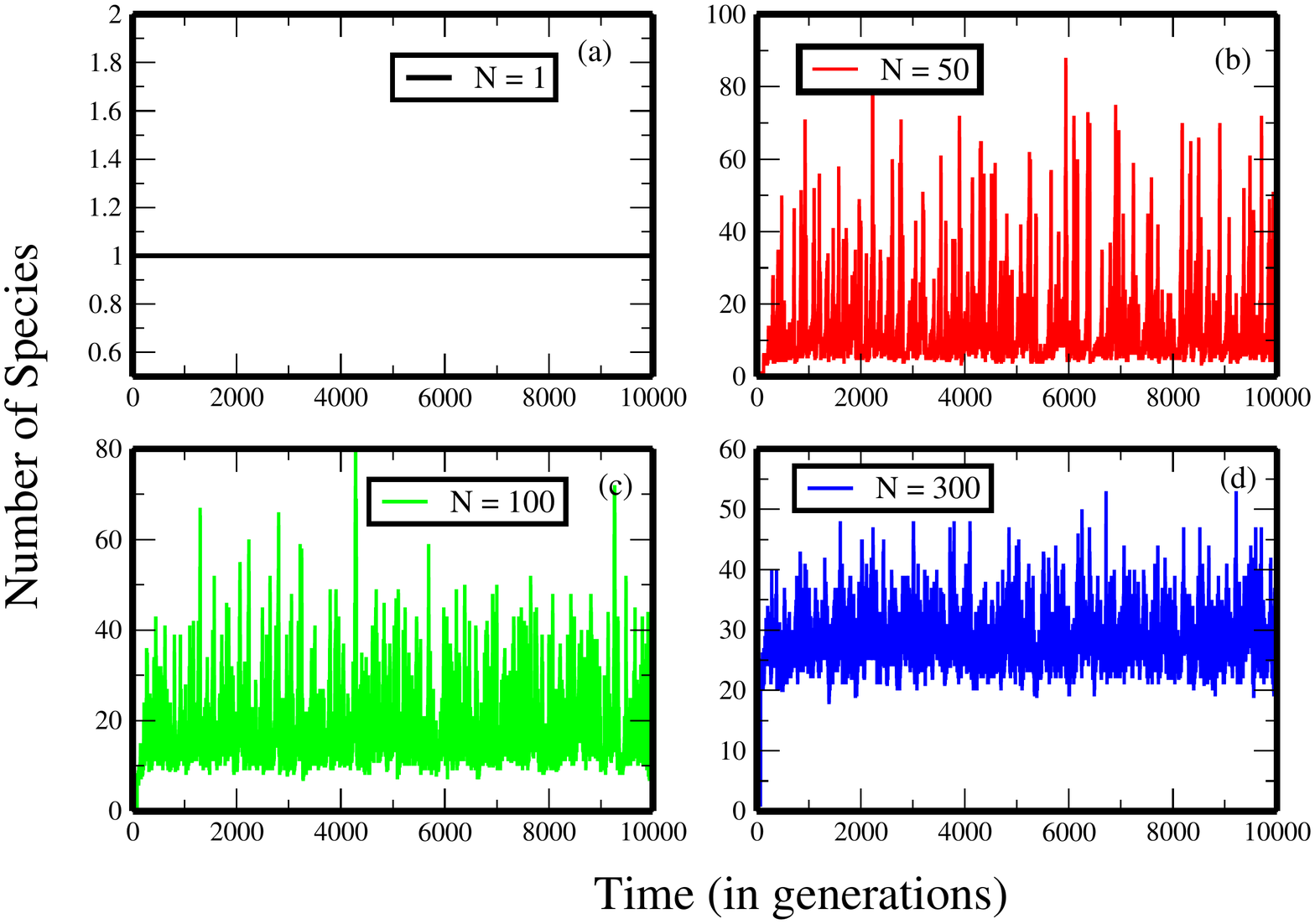}
 \caption{\label{fig:figure5} Time evolution of the number of species as a function of pool size for a population of $2000$ 
 	individuals and $\mu =1/8000$ for pool size equal to: (a) $N=1$, $q_0=0.504$, (b) $N=50$, $q_0=0.938$, (c) $N=100$, $q_0=0.96$ and 
 	(d) $N=300$, $q_0=0.966$. }
 \end{figure}

Figures \ref{fig:figure5} a,b,c and d show the number of species as a function of time for pool size
equal 1, 50, 100 and 300 individuals, respectively.  For each pool size the value of $q_0$ is
calculated. For small pool size ($N=1$, $q_0=0.504$, Figure \ref{fig:figure5}a)  there is only one
species. This case, where the choice of mates is completely random, is the same case
studied in \cite{higgs1991stochastic} where no minimal genetic similarity threshold is imposed and,
as it was anticipated, the number of species is one along the time and the diversity within the
specie is very high ($q_0\sim 0.5$). Nevertheless, once the formation of pair is not completely
random, one can see the emergence of new species as shown in Figures \ref{fig:figure5} b,c and d for
$N=50$ with $q_0=0.938$, $N=100$ with $q_0=0.96$ and $N=300$ with $q_0=0.966$, respectively.

\begin{figure}
\centering \includegraphics[width=15cm]{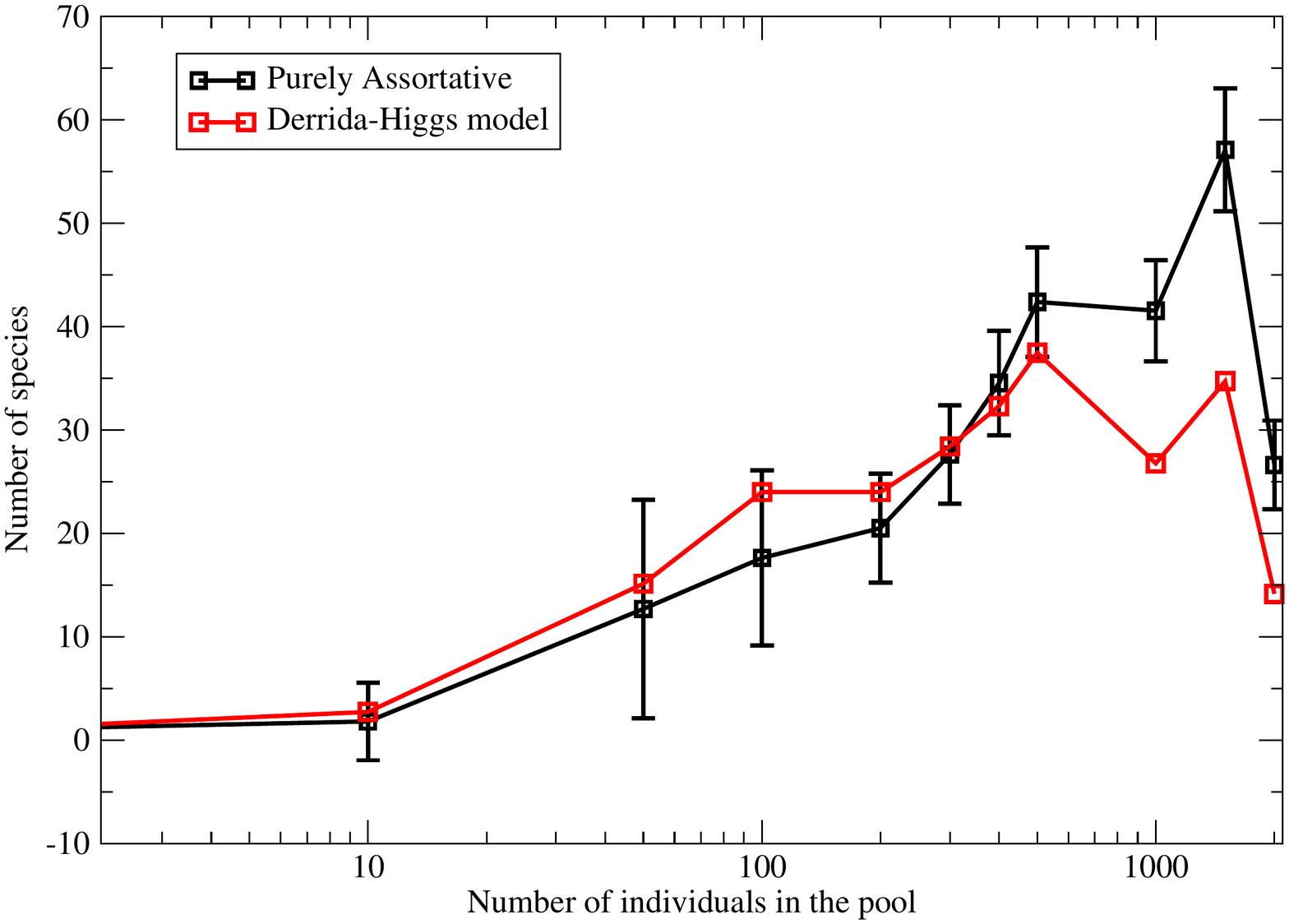} 
\caption{\label{fig:figure6} Numerical average
	number of species as a function of the pool size in a purely assortative reproduction is considered
	(black) and analytical average number of species for the DH model (red) for $M=2000$. Black dots represent
	average values computed at every generation and vertical bars are mean square deviations. The values of
$q_0$ for $N=1$, 10, 100 and 500  are 0.504, 0.732, 0.960 and 0.974 respectively.}
 \end{figure}

Figure \ref{fig:figure6} shows the time average and mean square deviation of the number of species
for different values of the pool size $N$ (black line) compared with the expected number of species
according to the DH model (red line). For each value of $N$ the parameter $q_0$ was calculated as
indicated above. The number of species shown is the average computed from the same time series at
every generation, from the initial time to 10,000. In the DH model the number of species
can be estimated by supposing that the population has $n$ species with, on average $m$
individuals, so that  $n m=M$. Using the fact that the mean value of the similarity of the group
with $m$ individuals drifts to the value $q(m)=1/(1+4\mu m)$ and, imposing $q(m)=q_{0}$,
we get $n=4\mu M(q_{0}^{-1}+1)^{-1}$.

\begin{figure}
\centering
\includegraphics[width=15cm]{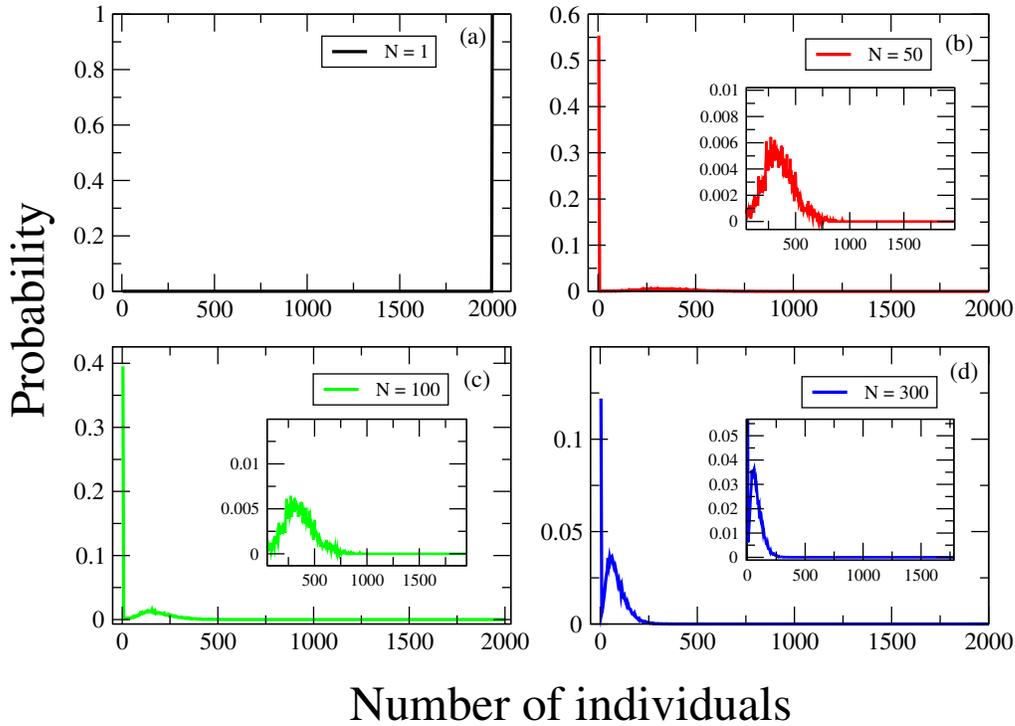}
 \caption{\label{fig:figure7} Abundance of a population with 2000 individuals after 10000 generations for pool sizes equal to 
 	a) 1, b) 50, c) 100 and d) 300 individuals. The abundance was calculated from the generations 3000 and 10000 in order 
 	to avoid the spurious effects of the transient time. The insets show a zoom of the distribution for large abundances.}
 \end{figure}

Finally, one important feature of a species is how many individuals it contains and how its size
depends on $N$, the number of individuals in the pool. Clearly, as the number of species varies
considerably, one could expect the same behavior for the number of individuals belonging to a
certain species. Thus, a useful quantity to measure this is the abundance, which has a probability
interpretation. The abundance is calculated as follows: after a specific transient time (we chose it
as the generation 3000), the number of individuals in each species is recorded for every generation.
At the end of the simulation (10000 generations), we construct a normalized histogram of the number
of occurrence of species with a given number of individuals. Figures \ref{fig:figure7} a,b,c and d
show the abundance for $N=1$, $N=50$, $N=100$ and $N=300$, respectively. $N=1$, shown in Figure
\ref{fig:figure7}a, is the trivial case in which there is only one species (see Figures
\ref{fig:figure5}a and \ref{fig:figure6}a) and this species contains all individuals. On the other
hand, as the pool size increases, a more complex probability distribution takes place. For example,
in Figure \ref{fig:figure7}b one can see a high probability of finding species with very few (or
even with one individuals) but a finite, although very low, probability of finding species with a
considerable number of elements. The peak at $1$ means that most of the individuals in the
population are less similar than the threshold $q_0$ to all the others (see also Figures
\ref{fig:figure8} and \ref{fig:figure9} below). For $N=100$, Figure \ref{fig:figure7}c,  the
probability of finding abundant species with around 150 individuals, is already significant. Figure
\ref{fig:figure5}d shows the case where the pool size is a substantial fraction of the system. In
such situation, although there exist species with only few elements, it is much more likely to find
robust species with a great number of individuals.
 
\subsection{Hybridization}

 An interesting situation that can be well described by our model is that of hybridization of two species. In  \cite{schumer2017} the authors reported the evolution of two species of swordtail ({\it Xiphophorus birchmanni} and {\it Xiphophorus malinche}) that maintain their reproductive isolation in sympatry because of assortative mating. Nevertheless, a disruption of pheromonal communication due to pollution lead to a breakdown of assortativeness \cite{Fisher} and to the formation of a hybrid population that have persisted for over 25 generations. 
	
\begin{figure}
		\centering
		\includegraphics[width=15cm]{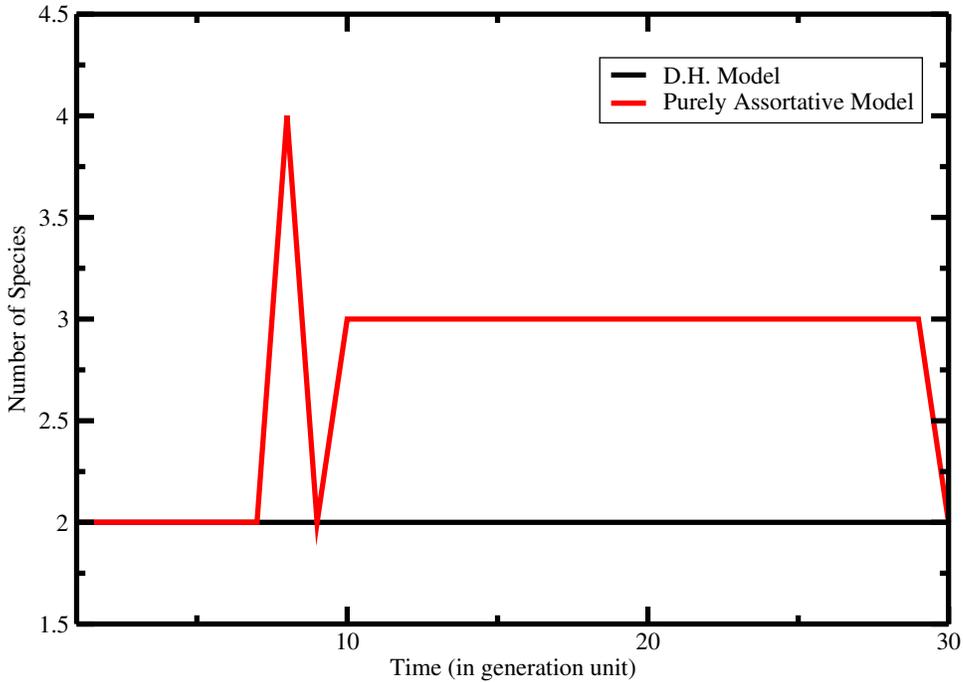}
		\caption{\label{fig:figure8} Time evolution of the number of species for the case where the assortativeness is broken at the first generation calculated by using our model (red line) and D.H. model (black line). After the first generation, the pool size is 400 individuals and the genetical similarity used to define specie is $q=0.97$. This $q$ is also used as genetic threshold in the D.H. model. }
\end{figure}

In order to show that the model can capture the essence of this process we simulated a population with two species (A and
B). Within each species, the individuals are initially genetically identical and the genetic similarity between
individuals belonging to different species was set to $0.9$. We simulated the evolution to the next generation without
assortativity (mimicking the effect of the pollution), allowing completely random mating (pool size equal 1). In the
next generation assortativity was restored with pool size of 400 individuals, simulating the recovery of the ecological
integrity of the environment. Figure \ref{fig:figure8} shows the time evolution of the number of species using
$q_{0}=0.97$ for species definition.
	
In order to compare the results of simulations with empirical observations we counted as species only groups with more than 10 individuals. The red line in Figure \ref{fig:figure8} shows the number of species along the time when the population evolves according to our model. The black line shows the time evolution according to the Derrida-Higgs dynamics. Since DH-model does not allow hybridization, the two initial groups are preserved and no new species appear during this initial stage of the dynamics. On the other hand, after a hybridization period (of around 7 generations), a new hybrid species appears which persists for about 20 generations.

\begin{figure}
	\centering
	\includegraphics[width=12cm,angle=270]{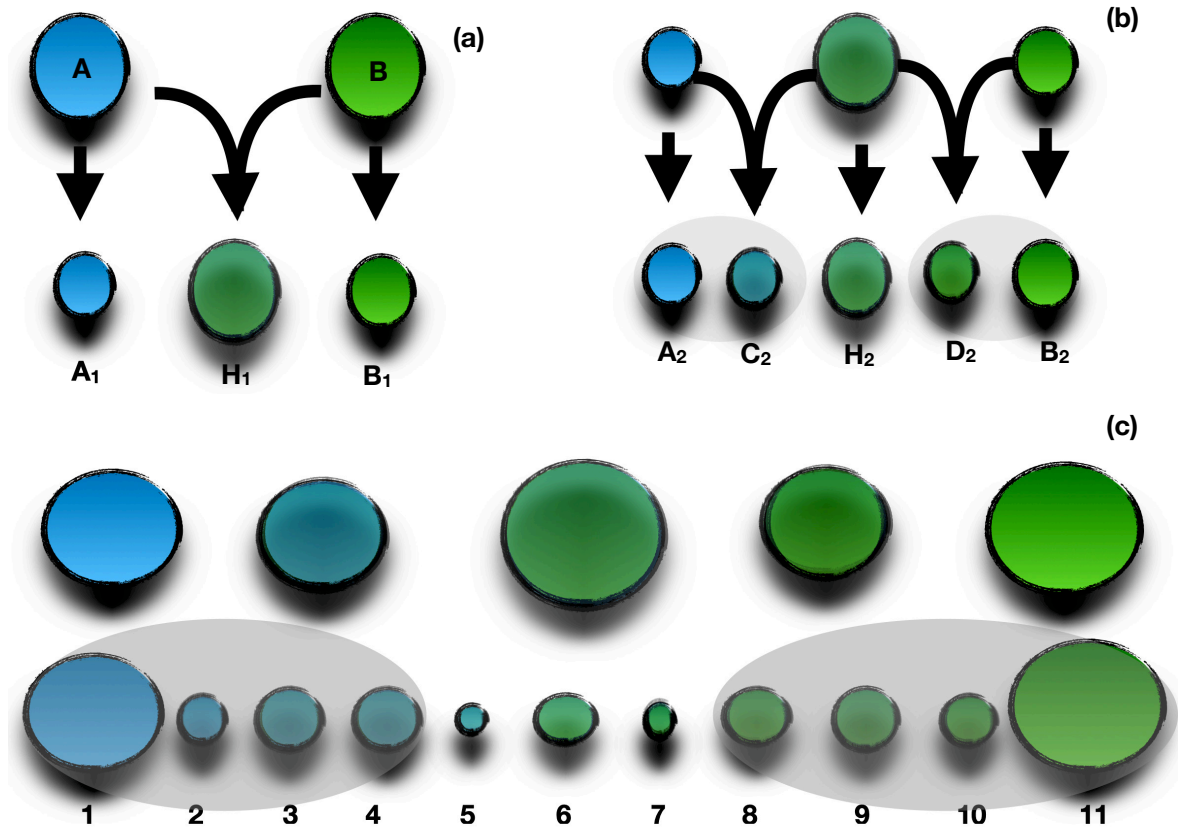}
	\caption{\label{fig:figure9} Illustrative scheme for the first (a), second (b) and third (c) generation of the population evolving according with our model.}
\end{figure}

The results of Figure \ref{fig:figure8} can be understood as follows: at generation 1 the population breaks into three groups. Two of those groups, $A_1$ and $B_1$ are `pure', meaning that their individuals are offspring of parents belonging to the same species, either A or B. Each of these groups has about $N/4$ individuals. The third group, $H_1$, with about $N/2$ individuals, contains the hybrids. The average genetic similarity between individuals belonging to $A_1$ (or $B_1$) is $q=e^{-4\mu}$ . The similarity between an individual in $A_1$ and one in $B_1$, on the other hand, is $q=0.9e^{-4\mu}$. Within the hybrid group $H_1$, or between $H_1$ and $A_1$ or $B_1$,  the genetic similarity is $q=0.95e^{-4\mu}$. If we set $q_0$ to any value below $0.95e^{-4\mu}$ these three groups would be identified as a single species, since individuals of $H_1$ would be able to mate with both $A_1$ and $B_1$, establishing gene flow among all individuals of the population. In order to identify $A_1$ and $B_1$ as different species we need to set $q_0$ above $0.9e^{-4\mu}$ and we  fixed  $q_0=0.97$. In this case only $A_1$ and $B_1$ are well defined species whereas the hybrid group $H_1$ forms a cluster whose individuals can still mate with any other if there is no mating threshold, but are too different from each other to be considered a species. The abundance distribution of the population would be similar to Figure \ref{fig:figure7}(b), with a peak at 1. This is illustrated in Figure \ref{fig:figure9}a. For the model with mating threshold at $0.97$ they will all die without living offspring and the population will maintain its two species for many generations as shown in Figure \ref{fig:figure8}. 

In generation 2 (and also thereafter) mating partners are selected from pools of size 400. Thus, if a first parent is selected from $A_1$ ($B_1$), the probability that the second parent is chosen from the hybrid group $H_1$ or from $B_1$ ($A_1$) is negligible. However, if a first parent is selected from $H_1$ the second parent can be taken from any of the three groups with significant probabilities. The second generation, therefore, has five groups, that we term $A_2$, $B_2$, $H_2$, $C_2$ and $D_2$, as illustrated in Figure \ref{fig:figure9}b). However, the genetic similarity between individuals from groups $C_2$ and $A_2$ and between individuals from groups $D_2$ and $B_2$ is $0.975e^{-8\mu}$ meaning that, according to our criterion of $q_0=0.97$, groups $C_2$ and $A_2$ and groups $D_2$ and $B_2$ belong to the same species. This is illustrated in Figure  \ref{fig:figure9}c.
	
In generation 3 there are still 2 robust species (with about 870 individuals each) and 1 hybrid group, as in Figure \ref{fig:figure9}c) (we have omitted the arrows indicating the origin of each group in order to make the illustration cleaner). Tracking the evolution of these subgroups along the following generations, one can observe that this pattern of more and more groups persists until a third species, with internal similarities larger than $q_0=0.97$, forms.  The  size of this hybrid species slowly decreases until its complete extinction in about 20 generations.

\section{Comparison with Empirical Data}

Although the model presented here discards any effects of spatial structure or selection, we have
attempted to compare our results with empirical data. In what follows we compare the results of our
simulations with a subgroup of African cichlids, the Haplochromines of Lake Victoria. Samonte {\it
	et al.} \cite{samonte2007gene} have analyzed the genetic variability of four species of the
haplochromine family collected from the Lake Victoria. The authors have compared the genetic
similarity of the four cichlid species in five {\it loci} of nuclear DNA. Here we used data from 
the {\it SINE1357 locus}, which has been used to study the phylogeny of salmon species \cite{murata1993determination} and
whales \cite{nikaido1999phylogenetic}.

\begin{figure}
	\centering
	\includegraphics[width=15cm]{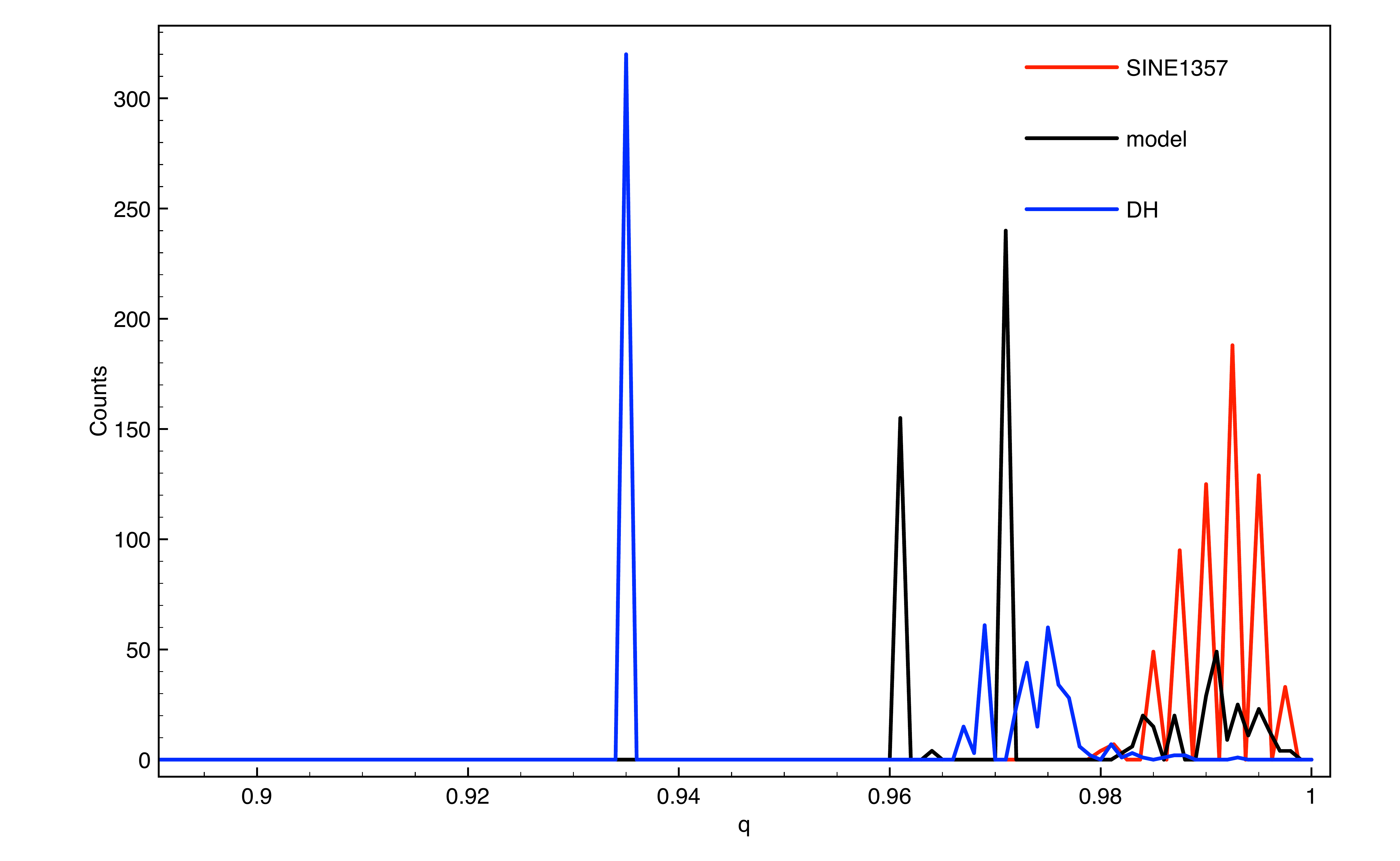}
	\caption{\label{fig:figure10} Similarity between 36 individuals of four species: Red line is the similarity of the {\it SINE1357 locus} of real individuals, the black line is the result obtained with our model and the blue line corresponds to the Derrida-Higgs model . }
\end{figure}

The authors have compared the genetic similarity of 36 individuals: 15, 16, 1 and 4 from {\it
	Haplochromis fischeri}, {\it  Ptyochromis xenognathus}, {\it Haplochromis chilotes} and {\it
	Haplochromis sp. ``rock kribensis"} species, respectively. The red line in Figure \ref{fig:figure10} corresponds to the histogram of similarity. It can be observed that there is a high degree of
similarity among all species. In order to compare with results from the model
considered here, we procedure as follow: First, we considered a population with 2000 individuals
with a pool size equals to 500 individuals, which gives a $q_{0}$ equals to 0.974. Second: from the
entire population, we select the four most similar (among them) species representing the four
species from \cite{samonte2007gene} and, third, we take, randomly, 15, 16, 1 and 4 individuals from
these species. The histogram of these 36 individuals is shown by the black line in Figure
\ref{fig:figure10}. It is noticeable that the similarity between many individuals is very consistent
with the similarity of the individuals from \cite{samonte2007gene}. The main difference with
experimental data is the two peaks around $q=0.96$ and $q=0.97$. We believe the main reason for such
difference is that we have compared the infinite genome of the theoretical population while only a
small fraction of the genome in the real population. We also compared the similarity between
individuals from a theoretical population that evolved according to the Derrida-Higgs model
\cite{higgs1991stochastic} using $q_{min}=q_{0}=0.974$, shown by the blue line in Figure
\ref{fig:figure10}. One can see that only the threshold for genetic similarity is not enough to
promote the good qualitative agreement as the assortative model and the experimental data.

 \section{Discussion}
 
 In this work we have considered a variation of the DH model in which the genetic threshold for reproduction
compatibility is completely eliminated and replaced by a pure assortative mating mechanism.
In our model each individual chooses for mating the most similar partner it can find in a random pool of size $N$, no
matter how dissimilar it might be. Figure \ref{fig:figure2} shows that if the first parent can choose its mate from a
pool of individuals of sufficiently large size, the histogram of similarities $q$'s  does exhibit peaks below the
asymptotic value $q_{conv}$. However, for small pools, a different dynamics take place, as illustrated in Figure
\ref{fig:figure3} for $N=10$ (left panels). In these cases the distribution might break into clusters centered at
$q>q_{conv}$ but the inter-cluster similarity is still large. These clusters are genetically close to each other and
unstable. Since the pool is too small the likelihood that an individual of a cluster will select a pool consisting of
individuals belonging solely to other clusters is high, making the clusters merge by the introduction of hybrids. As
pool size increases the genetic distance between the clusters also increases and the peak representing inter-cluster
similarity moves to the left of $q_{conv}$. For $N=30$, the clusters are still unstable and species prone to merging
with one another, as illustrated in Figure \ref{fig:figure3} (right panels). For large pools the probability that the
pool contains individuals of the same cluster is large, maintaining the species cohesion. However, merging is always
possible since reproduction between any pair produces viable offspring. Thus, mating between individuals from different
clusters does take place, generating hybrids and potentially merging species.
 
 Our results demonstrate that assortative mating can lead to formation of clusters of individuals with high degree of
similarity among themselves but with low degree of similarity with individuals belonging to other clusters. The number
of species in the population, however, fluctuates in time, as shown in Figure \ref{fig:figure6}. For small and medium
pool sizes, there is a clear tendency of increasing the number of species as the pool size increases. However, for pool
sizes larger than approximately 15-20\% of the population, the behavior is more complex. In this regime the number of
species  is more sensitive to the reference value $q_{0}$. This fact can be understood as follows: if the pool size $N$
is a considerable fraction of the total population, the probability that the individual with the highest similarity
with the first parent is in this pool is $N/M$, which is large  by hypothesis. Moreover, even if the individual with
highest similarity is not present in the pool, it is very likely that the highest similarity between the first parent
and the individuals from the pool is very close to the highest similarity between the first parent and an individual
from the population.
 
 It is important to compare the model presented in this work with the one introduced in \cite{higgs1991stochastic}
where a minimum similarity is required in order to produce offspring (Figure \ref{fig:figure6}). Within the error bars,
both models provide almost the same number of species for small and medium pool size. On the other hand, when the pool
size is a considerable fraction of the population, the purely assortative model leads to a larger number of species.
This behavior is a consequence of the fact that, for large pool sizes, assortative mating can be more restrictive than
the constraint of minimum similarity.  On average, the similarity between parents in DH model is smaller than the
similarity of the parents in the assortative model when the pool size is a considerable fraction of the population.
 
Finally, we have compared our model with data from a subgroup of cichlids from Victoria lake. Although our model is
 very simple and discards important features of the real dynamics (such as spatial structure, finite character of the real
 genomes and other selective forces), we obtained qualitative good agreement with the data, slightly better than that
 provided by the Derrida-Higgs model with a genetic threshold.

To summarize, we have introduced an evolutionary model based purely on assortative mating, without a
genetic threshold that impairs reproduction. In the model individuals mate with the most similar
individual in their pool of potential partners. We have shown that this condition alone can lead to
the formation of clusters of genetically similar individuals. Although the absence of reproductive
isolation makes the definition of species dependent on the identification of these clusters, which
has a degree of  subjectivity, these results show that species can potentially emerge from sympatric
condition and without the need of an arbitrary genetic threshold to create reproductive isolation.
We have studied the formation and evolution of a hybrid species formed  by mixing two isolated
species during a single generation of environmental disturbance. Restoring the environmental
integrity in the next generation results in the formation of a hybrid species that slowly decreases
in abundance and finally goes extinct, as observed in empirical studies.

\vspace{1cm}

\noindent Acknowledgments: M.A.M.A.  acknowledges financial support from CNPq (grant 302049/2015-0) and
FAPESP (grants 2016/06054-3 and 2015/11985-3). C.L.N. Costa. was supported by Capes, grant 302049/2015-0.

\providecommand{\newblock}{}

\end{document}